\documentclass[12pt]{iopart}

\usepackage{iopams}
\usepackage{upgreek}
\usepackage{graphicx}
\usepackage{amssymb,amsfonts}

\newcommand{\half}{\mbox{$\textstyle \frac{1}{2}$}}
\def\opone{\leavevmode\hbox{\small1\kern-3.8pt\normalsize1}}
\newcommand{\unit}[1]{\ensuremath{\,\mathrm{#1}}} 
\newcommand{\degc}{\ensuremath{\,^\circ\mbox{C}}}

\begin{document}


\title{Characterizing heralded single-photon sources with imperfect measurement devices}

\author{M Razavi${}^{1,2,3}$, I S\"ollner${}^{1,4}$, E Bocquillon${}^{1,5}$, C Couteau${}^{1,3}$, R Laflamme${}^{1,3,6}$ and G Weihs${}^{1,4}$}

\address{${}^1$Institute for Quantum Computing, ${}^2$Department of Electrical and Computer Engineering, and ${}^3$Department of Physics and Astronomy, University of Waterloo, 200 University Ave. W., Waterloo, ON, Canada  N2L 3G1}
\address{${}^4$Institut f\"ur Experimentalphysik, Universit\"at Innsbruck, Technikerstrasse 25, 6020 Innsbruck, Austria}
\address{${}^5$Ecole Normale Sup\'{e}rieure, 45, rue d'Ulm, 75230 Paris, France}
\address{${}^6$Perimeter Institute for Theoretical Physics, 31 Caroline St., Waterloo, ON, Canada N2L 2Y5}

\ead{mrazavi@iqc.ca}

\begin{abstract}
Any characterization of a single-photon source is not complete without specifying its second-order degree of coherence, i.e., its $g^{(2)}$ function. An accurate measurement of such coherence functions commonly requires high-precision single-photon detectors, in whose absence, only time-averaged measurements are possible. It is not clear, however, how the resulting time-averaged quantities can be used to properly characterize the source. In this paper, we investigate this issue for a heralded source of single photons that relies on continuous-wave parametric down-conversion. By accounting for major shortcomings of the source and the detectors---i.e., the multiple-photon emissions of the source, the time resolution of photodetectors, and our chosen width of coincidence window---our theory enables us to infer the true source properties from imperfect measurements. Our theoretical results are corroborated by an experimental demonstration using a PPKTP crystal pumped by a blue laser, that results in a single-photon generation rate about 1.2 millions per second per milliwatt of pump power. This work takes an important step toward the standardization of such heralded single-photon sources.
\end{abstract}

\pacs{42.50.Dv, 42.50.Ar, 42.65.Lm, 03.67.Dd}

\maketitle

\section{Introduction}

Single-photon sources (SPSs) are important elements in quantum communication, optical quantum computing, and metrology \cite{Bennett84a}. To satisfy the requirements for such applications, it is desired that such SPSs only create single, and {\em not} multiple, photons in an {\em on-demand} way. To build such a source, one can employ an array of {\em heralded} single-photon sources (HSPS) \cite{HSPS}, with possibly low individual probability of single-photon generation, but with an overall probability that approaches one for a sufficiently large number of sources in the array \cite{Shapiro07a}. One of the most convenient ways to generate heralded single photons is based on spontaneous parametric down-conversion (SPDC) \cite{Hong86}. In this scheme, by pumping a crystal with $\chi^{(2)}$ nonlinearity, one can generate a pair of signal and idler beams whose photon numbers in given time intervals are highly correlated, thus by detecting a single photon on the idler beam one can, ideally, guarantee the presence of a single photon on the signal beam. Inevitable to this scheme is the occasional generation of multiple-photon packets in each beam, which degrade the reliability of the SPS. This effect can be best examined quantitatively by evaluating the degree of second-order coherence (termed coherence function, hereafter, for brevity) for our source. In this paper, we analytically calculate the coherence function for our SPDC-based source and measure it in an experiment. We consider the impact of finite time resolutions on the measurement results, and how this shortcoming may affect the proper characterization of such sources. In fact, our work paves the way for developing standard specifications for HSPSs.

For an ideal SPS, we expect that its second-order degree of coherence, viz. its $g^{(2)}$ function, is zero at the origin \cite{Loudon83a}. This is equivalent to having no coincidence detection on the two detectors of a Hanbury-Brown and Twiss (HBT) interferometer \cite{Brown56a}. ``Coincidence,'' in theory, refers to two simultaneous events. In practice, however, a coincidence event can only be verified within a certain accuracy permitted by the employed measurement devices, e.g. photodetectors and their respective electronics. This requires us to define coincidence by referring to two events that occur within a coincidence window whose width is greater than zero. The measured values for $g^{(2)}$---or even its simplified form commonly expressed as the probability of having two or more photons in a given (short) time interval over the probability square of having only one photon---may well depend on our choice of coincidence window as well as on other experimental parameters. Such dependence poses a challenge on the proper standardization of HSPSs because a single value of $g^{(2)}(0)$ does not necessarily convey sufficient information to characterize such a source. This is particularly the case in our continuous-wave (cw) SPDC-based HSPS, whose coherence functions may have widths in the sub-picosecond regime, much lower than what typical photodetectors can measure.

The measurement of coherence functions in SPDC-based HSPSs is not only affected by the above time parameters but also by the multiple-photon contribution to the SPDC output. The latter is a function of the pump power, which, at the same time, determines the rate of single-photon generation of our source. In this paper, we present a theoretical framework that not only, for the first time, accounts for the multiple-photon emission in our source but also allows us to examine the effect of imperfect devices on the $g^{(2)}$ measurement. Such an analysis provides prescriptions for proper characterization of coherence properties of HSPSs and how such figures can be measured in practical experimental setups. This is of crucial importance because such devices have already been introduced into the market \cite{market}. We accompany our theoretical work with experimental evidence using a collinear setup for our type II periodically poled KTP (PPKTP) crystal. Our theory is well capable of reproducing the measurement results.

The rest of the paper is organized as follows. In Sec.~\ref{Sec_Theory}, we develop the theoretical model for our HSPS, and evaluate its second-order coherence properties as functions of source parameters in the ideal limit of infinitely high time resolutions. For this purpose, we use a heuristic continuous-mode analysis, whose validity is confirmed by an asymptotic discrete-mode analysis presented in the Appendix. Section~\ref{Sec_setup} describes the experimental setup for our HSPS and the corresponding HBT interferometer, followed by our experimental results in Sec.~\ref{Sec_results}. There, we introduce our time-averaged coherence functions and their relation to the ideal figures. Section~\ref{Sec_Conclusion} concludes the paper.

\section{Theory of SPDC-based Single-Photon Sources}
\label{Sec_Theory}

The HSPS considered here consists of a parametric down-converter---driven by a cw pump at center frequency $\omega_p$ producing cw signal ($s$) and idler ($i$) beams at center frequencies $\omega_s$ and $\omega_i = \omega_p -\omega_s$, respectively---followed by a single-photon detector on the idler beam. Here, we implicitly assume that signal and idler beams can be separated into two orthogonal spatial modes. In our experimental setup, this has been achieved by employing a type-II crystal, which creates signal and idler beams with orthogonal polarizations, along with a polarizing beam-splitter (PBS). Here, for simplicity, we suppress the spatial and polarization characteristics of signal and idler beams and represent them with scalar photon-units positive-frequency field operators \cite{Shapiro85a}:
\begin{equation}
\label{Eq:field_op}
\hat E_j(t) = \frac{1}{{2 \pi}} \int {d \omega \hat A_j(\omega) e^{-i \omega t} }, \quad\mbox{$j=s,i$,}
\end{equation}
where $[\hat E_j(t),\hat E_j^\dag(u)]=\delta(t-u)$ and $\hat A_j(\omega)$ represents the corresponding output field operator in the frequency domain. It has been shown that, in the Heisenberg picture, the output field operators can be related to the vacuum-state field operators at the input to the crystal, $\hat A_j^{in}(\omega)$, $j=s,i$, via the following Bogoliubov transformation \cite{Shapiro94a}
\begin{eqnarray}
\label{Eq:Bogo1}
\hat A_s(\omega_s + \omega)& = & \mu(\omega) \hat A_s^{in}(\omega_s + \omega) + \nu(\omega) \hat A_i^{in \dag}(\omega_i - \omega),\\
\label{Eq:Bogo2}
\hat A_i(\omega_i - \omega)& = & \mu(\omega) \hat A_i^{in}(\omega_i - \omega) + \nu(\omega) \hat A_s^{in \dag}(\omega_s + \omega),
\end{eqnarray}
where $|\mu(\omega)|^2-|\nu(\omega)|^2 = 1$. The joint state of signal and idler is a zero-mean Gaussian state whose only nonzero second-order moments are given by its temporal auto- and cross-correlation functions as follows \cite{Shapiro94a,Wong06a}
\begin{eqnarray}
\label{Eq:autocor}
&\!\!\!\!\!\!\!\!\!\!  \langle \hat E_j^\dag(t+\tau) \hat E_j(t) \rangle = e^{i \omega_j \tau} R(\tau), \,\, R(\tau) \equiv \int{ \frac {d \omega}{2 \pi}|\nu(\omega)|^2 e^{i \omega \tau} }, &\\
\label{Eq:crosscor}
&\!\!\!\!\!\!\!\!\!\! \langle \hat E_j(t+\tau) \hat E_k(t) \rangle=  (1-\delta_{jk})e^{-i \omega_p t - i \omega_j \tau}  C(\tau), \,\, C(\tau) \equiv \int{ \frac {d \omega}{2 \pi}\nu(\omega) \mu(\omega) e^{-i \omega \tau} }, &
\end{eqnarray}
where $\delta_{jk}$ is the Kronecker delta function and $j,k=s,i$.

In the low-gain regime, which is of interest to us, $\mu(\omega) \approx 1$ and $|\nu(\omega)|^2 \approx (R_{\rm SPDC} / B_{\rm SPDC}) \sin^2[\omega/(2B_{\rm SPDC})]/[\omega/(2B_{\rm SPDC})]^2$, where $R_{\rm SPDC}$ is the rate of photon generation for the signal/idler beam, and $B_{\rm SPDC}$ is the bandwidth of the SPDC process \cite{Wong06a}. In this regime, we have
\begin{equation}
\label{Rcorr}
R(\tau) = \left\{\begin{array}{cc}
R_{\rm SPDC} (1+ \tau B_{\rm SPDC} ) & -\frac{1}{B_{\rm SPDC}} < \tau \leq 0 \\
R_{\rm SPDC} (1- \tau B_{\rm SPDC}) & 0 < \tau \leq \frac{1}{B_{\rm SPDC}} \\
0 & \mbox{elsewhere}
\end{array} \right. ,
\end{equation}
and
\begin{equation}
\label{Ccorr}
|C(\tau)| = \left\{\begin{array}{cc}
\sqrt{R_{\rm SPDC} B_{\rm SPDC}}  & -\frac{1}{2 B_{\rm SPDC}} < \tau < \frac{1}{2 B_{\rm SPDC}} \\
0 & \mbox{elsewhere}
\end{array} \right. ,
\end{equation}
where we assume that the difference in the speed of light for ordinary and extraordinary axes in the crystal has been compensated. In our experiment, $B_{\rm SPDC}$ is on the order of THz, and $R_{\rm SPDC}$ is on the order of 1\unit{MHz} per milliwatt of pump power.

In this paper, we calculate two coherence measures for our HSPS. The first figure quantifies the reliability of our heralding mechanism by looking at the temporal correlation between the signal and idler beams, and the second measure quantifies its capability to create one---and only one---photon per heralding event. In both cases, we first find the full temporal shapes of the above coherence functions in the ideal limit of infinitely high time resolutions and, then, later in Sec.~\ref{Sec_results} we will introduce our corresponding time-averaged parameters that we can measure in a typical experimental setup. Also, throughout the paper, we neglect the dark count effect and we assume that all employed photodetectors have unity quantum efficiencies. The latter assumption does not affect our measurement results because all correlation functions that we deal with in this paper have normalized forms.

\subsection{Signal-idler Temporal Correlation}

As a measure of temporal correlation between signal and idler, we obtain the degree of  second-order coherence between the signal and the idler fields defined as follows,
\begin{eqnarray}
\label{Eq:g_si_2}
g_{si}^{(2)}(t+\tau,t) & \equiv & \frac {\langle \hat E_s^\dag(t+\tau) \hat E_i^\dag(t) \hat E_i(t) \hat E_s(t+\tau) \rangle} {\langle \hat E_s^\dag(t+\tau) \hat E_s(t+\tau) \rangle \langle \hat E_i^\dag(t) \hat E_i(t) \rangle} \nonumber \\
& = & 1 + \frac{|C(\tau)|^2} {R^2(0)} \equiv g_{si}^{(2)}(\tau),
\end{eqnarray}
where, in the last step, we used the quantum form of the Gaussian moment-factoring theorem  \cite{Shapiro94a} by which we can reduce the
fourth-order moment in the above equation to the sum of products of second-order moments, available from Eqs.~(\ref{Eq:autocor}) and
(\ref{Eq:crosscor}), as follows
\begin{eqnarray}
\label{Psit1t2}
P_{si}(t+\tau,t) & \equiv & {\langle \hat E_s^\dag(t+\tau) \hat E_i^\dag(t) \hat E_i(t) \hat E_s(t+\tau) \rangle} \nonumber \\
& = & {\langle \hat E_s^\dag(t+\tau) \hat E_i^\dag(t) \rangle \langle \hat E_i(t) \hat  E_s(t+\tau) \rangle} \nonumber \\
& + &  {\langle \hat E_s^\dag(t+\tau) \hat E_i(t) \rangle \langle \hat E_i^\dag(t)  \hat  E_s(t+\tau) \rangle} \nonumber \\
& + & {\langle \hat E_s^\dag(t+\tau)  \hat  E_s(t+\tau) \rangle \langle \hat E_i^\dag(t) \hat E_i(t)  \rangle} \nonumber \\
& = & R^2(0) + |C(\tau)|^2 \equiv P_{si}(\tau).
\end{eqnarray}
Here, $P_{si}(t+\tau,t)$ is the coincidence rate for observing a signal photon at time $t+\tau$ and an idler photon at time $t$, and from the above equation, it is only a function of $\tau$. In the low-gain regime, $g_{si}^{(2)}(0) \approx B_{\rm SPDC}/R_{\rm SPDC}$, which is inversely proportional to the probability of detecting
a photon in a time interval of width $\Delta {\rm t} \equiv 1/B_{\rm SPDC}$. For our experimental setup, $g_{si}^{(2)}(0)$ is on the order of $10^5$ and $g_{si}^{(2)}(\tau)$ has a narrow sub-picosecond width. These two properties witness ultrashort, highly correlated, twin wavepackets.

\subsection{Second-order Coherence Function for Heralded Signal}

The second coherence measure that we consider here is the degree of second-order coherence for the signal field, conditioned on observing an idler photocount at time $t_i$, defined as follows
\begin{equation}
\label{Eq:Cond_g2}
g_c^{(2)}(t_1,t_2|t_i) \equiv \frac{\langle \hat E_s^\dag(t_1) \hat E_s^\dag(t_2) \hat E_s(t_2) \hat E_s(t_1) \rangle_{\rm pm}}{\langle \hat E_s^\dag(t_1) \hat E_s(t_1) \rangle_{\rm pm}  \langle \hat E_s^\dag(t_2) \hat E_s(t_2) \rangle_{\rm pm}},
\end{equation}
where $\langle \cdot \rangle_{\rm pm}$ is the average over the post-measurement state assuming sufficiently high time resolution and unity quantum efficiency for the idler photodetector.

To model the measurement on the idler field operator, we use a heuristic approach in which a photodetection event at time $t_i$ on the idler beam is modeled by the continuous-time measurement operator, \cite{Nielsen00a}, $\hat E_i(t_i)$. In the Appendix, we employ a discrete-mode formalism for the same problem and show that in the asymptotic limit of infinitely high time resolution the results of the two methods converge. The post-measurement averaging, for any operator $\hat X$, will then be given by
\begin{equation}
\langle \hat X \rangle_{\rm pm} = {\langle \hat E_i^\dag(t_i) \hat X \hat E_i(t_i)\rangle} / { {\langle  \hat E_i^\dag(t_i)\hat E_i(t_i)\rangle}} .
\end{equation}
The conditional coherence function in Eq.~(\ref{Eq:Cond_g2}) can then be written as follows
\begin{equation}
\label{Eq:gct1t2}
g^{(2)}_c(t_1,t_2|t_i)=\frac{P_{si}^{(2)}(t_1,t_2,t_i) R(0)}{P_{si}(t_1,t_i)P_{si}(t_2,t_i)},
\end{equation}
where, using again the quantum version of the Gaussian moment-factoring theorem along with Eqs.~(\ref{Eq:autocor}) and (\ref{Eq:crosscor}),
\begin{eqnarray}
\label{Psit1t2ti}
P_{si}^{(2)}(t_1,t_2,t_i) & \equiv & \langle \hat{E}_i^\dag(t_i)\hat{E}_{s}^\dag(t_1)\hat{E}_{s}^\dag(t_2)\hat{E}_{s}(t_2)\hat{E}_{s}(t_1)
\hat{E}_i(t_i) \rangle \nonumber \\
& = & R(0)\left[R^2(0)+|R(\tau_{12})|^2 + |C(\tau_{1})|^2 + |C(\tau_{2})|^2 \right] \nonumber \\
&\,& + 2\Re \left\{C(\tau_{1})C^\ast(\tau_{2})R(\tau_{12})\right\}
\end{eqnarray}
is the multi-coincidence rate for finding signal photons at times $t_1$ and $t_2$ and an idler photon at time $t_i$. In the above equation,
$\tau_{12} \equiv t_1-t_2$, $\tau_{1} \equiv t_1-t_i$, and $\tau_{2} \equiv t_2-t_i$. Finally, by plugging Eqs.~(\ref{Psit1t2ti}) and (\ref{Psit1t2})
into Eq.~(\ref{Eq:gct1t2}), we find
\begin{eqnarray}
\label{Eq:gct1t2ti}
g^{(2)}_c(t_1,t_2|t_i)& = & \frac {1}{g_{si}^{(2)}(\tau_1)} + \frac {1}{g_{si}^{(2)}(\tau_2)}  + \frac{|R(\tau_{12})|^2/R^2(0)-1}{g_{si}^{(2)}(\tau_1)g_{si}^{(2)}(\tau_2)} \nonumber \\
& + & \frac{2 \Re \left\{C(\tau_{1})C^\ast(\tau_{2})R(\tau_{12})\right\}}{R^3(0) g_{si}^{(2)}(\tau_1)g_{si}^{(2)}(\tau_2)}.
\end{eqnarray}

There are several interesting cases to be considered. First, let us look at the coherence function at the trigger time, i.e.,
\begin{equation}
\label{gc20}
g^{(2)}_c(t_i,t_i|t_i) = \frac{2}{g_{si}^{(2)}(0)} \left(2- \frac{1}{g_{si}^{(2)}(0)} \right).
\end{equation}
It is clear that if $g_{si}^{(2)}(0) \gg 1$ then
$g^{(2)}_c(t_i,t_i|t_i) \approx 0$ as desired. In other words, the reliability of the heralding mechanism as well as the multiple-photon suppression are both guaranteed by the same condition $R^2(0) \ll |C(0)|^2$.

The second interesting case is when $t_1 = t_i$ but $|\tau_2| = |t_2-t_i| \gg 2 \Delta\mathrm t$. In this case,
\begin{eqnarray}
g^{(2)}_c(t_i,t_2|t_i)& = & \frac {1}{g_{si}^{(2)}(0)} + \frac {1}{g_{si}^{(2)}(\tau_2)} + \frac{|R(-\tau_{2})|^2/R^2(0)-1}{g_{si}^{(2)}(0)g_{si}^{(2)}(\tau_2)}
\nonumber \\
& + & \frac{2 \Re \left\{C(0)C^\ast(\tau_{2})R(-\tau_{2})\right\}}
{R^3(0)g_{si}^{(2)}(0)g_{si}^{(2)}(\tau_2)} \nonumber \\
&\approx& 1,
\end{eqnarray}
provided that $g_{si}^{(2)}(0) \gg 1$ and $ g_{si}^{(2)}(\tau_2)  \approx 1$. This implies that our HSPS has a coherence time on the order of $\Delta\mathrm t$.

Finally, let us consider the case when $|\tau_1 = \tau_2| \gg 2 \Delta\mathrm t$, i.e, when there is no correlation between the trigger time and the signal beam. In this case,
\begin{eqnarray}
g^{(2)}_c(t_2,t_2|t_i)& = & 2 + \frac{2}{g_{si}^{(2)}(\tau_2)} \left(1- \frac{1}{g_{si}^{(2)}(\tau_2)} \right) \nonumber \\
&\approx& 2,
\end{eqnarray}
provided that $ g_{si}^{(2)}(\tau_2)  \approx 1$, which prevails in the low-gain regime. This is in accord with the fact that in the SPDC process, in the lack of any triggering event, both signal and idler beams individually obey the thermal-state statistics, for which the second-order coherence function has a maximum value two \cite{Loudon83a}.

\section{Experimental Setup}
\label{Sec_setup}

\begin{figure}
    \begin{center}
        \includegraphics[width=0.6\textwidth]{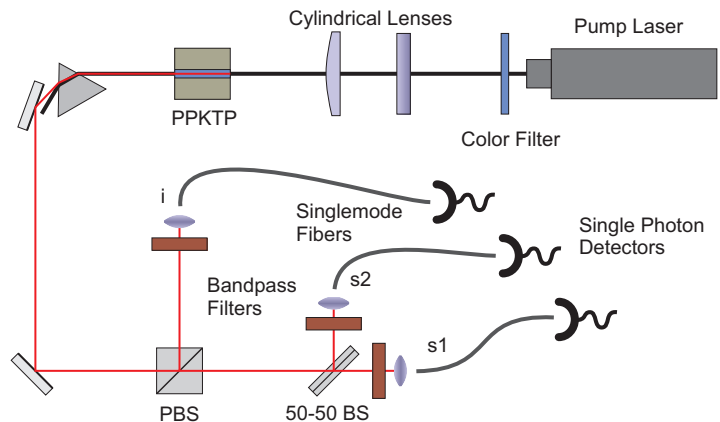}
    \end{center}
    \caption{Experimental setup for our heralded single-photon source. A blue laser pumps a $1\times2\times10\unit{mm^3}$ PPKTP crystal to create signal and idler beams. The pump beam will be removed by using dichroic filters as well as interference filters. Signal and idler beams are split into different spatial modes by using a polarizing beam splitter (PBS). The idler beam is used as a trigger and the signal beam goes through a 50/50 beam splitter for the $g^{(2)}$ measurement.}
    \label{Fig:setup}
\end{figure}

\begin{figure}
    \begin{center}
        \includegraphics[width=0.6\textwidth]{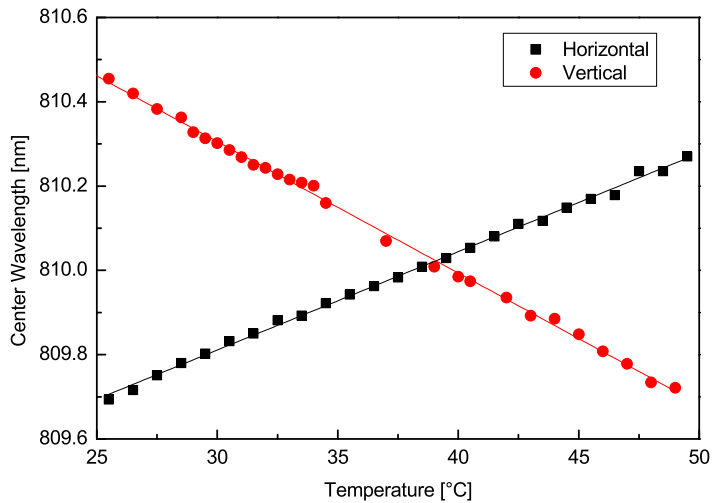}
    \end{center}
    \caption{Tuning curve of the signal (Horizontal) and idler (Vertical) photon center wavelengths as a function of the crystal temperature.}
    \label{Fig:temp}
\end{figure}

In this section, we describe the experimental setup used to demonstrate the theoretical findings from the previous section. Figure~\ref{Fig:setup} presents the optical setup used for our HSPS along with the HBT interferometer used for the $g_c^{(2)}$ measurement. A cw blue laser at center wavelength 405\unit{nm}  pumps a type-II periodically-poled $\mathrm{KTiOPO_4}$ (PPKTP) crystal. The crystal was from Raicol with a 10\unit{\upmu m} period and its dimensions were $1\times2\times10\unit{mm^3}$. The periodicity was chosen so that we would achieve creation of degenerate photon pairs at 810~nm slightly above the room temperature. The PPTKP crystal was inserted into a home-made oven made from copper and surrounded by PEEK plastic, which allowed us to reach the degeneracy at $39\degc$, as shown in Fig.~\ref{Fig:temp}, with a stability of $\pm 0.1\degc$. Some optical elements were used to focus the laser, to reshape its spatial mode, and to collect the signal and idler beams. Dichroic filters were employed after the crystal to remove the pump beam. The crystal was cut for propagation along the x-axis in order to support mainly type-II SPDC for which the signal and idler photons have orthogonal polarizations. With the help of a PBS we could then deterministically split the two beams into two different spatial modes. A photodetection event on the idler beam heralds the presence of one or more photons on the signal beam, which goes through an HBT interferometer consisting of a 50/50 beam splitter, two interference filters, and two single-photon photodetectors. All photodetectors were single-photon counting modules from Perkin-Elmer with equal nominal quantum efficiencies of 0.4, dead-times of 45\unit{ns}, and time resolutions of 350\unit{ps}. Our interference filters had a 10\unit{nm} bandwidth centered at wavelength 810\unit{nm}. The measured photon count rate for our setup was up to around 850,000 counts/s in each channel at 50$\,$mW pump power, with a signal-idler coincidence count rate amounting to at most about 10\% of that value, which implies that the mode-matching was still far from perfect. Moreover, in order to avoid overloading the data acquisition hardware we occasionally chose to attenuate the down-converted beams with neutral density filters, which reduced the coincidence count rate even further without affecting the coherence properties of the source.

The detection times for the signal and idler beams were recorded by a time-tagging card from Dotfast Consulting with a nominal temporal resolution of 156.25\unit{ps}. The time-tagging card streams the time tags to a computer by which we could calculate any single, double, or triple coincidence rates between the three channels ($i$, $s_1$, and $s_2$) in Fig.~\ref{Fig:setup} with a coincidence window that could be varied from 0.5\unit{ns} to 20\unit{ns}. The complete system of photodetectors, power supplies, time-stamping electronics and the USB interface fits in a $30\times30\times30\unit{cm^3}$ box.

\begin{figure}
    \begin{center}
        \includegraphics[width=0.6\textwidth]{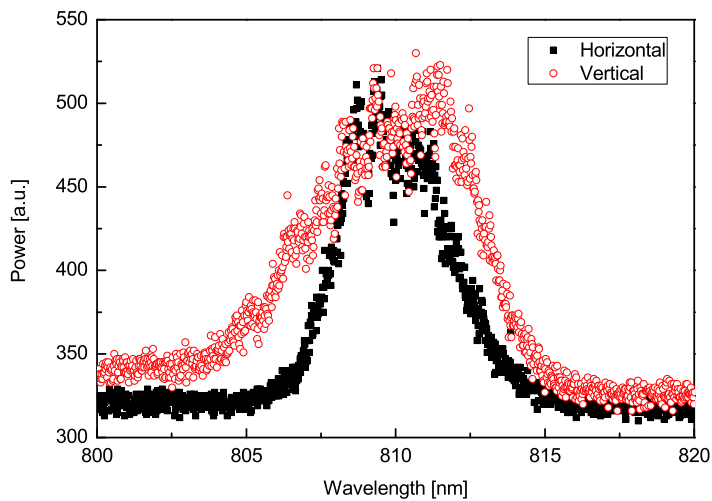}
    \end{center}
    \caption{Spectrum of the down-converted pair of photons at nominal degeneracy at 39 °C with center-of-mass wavelengths of 809.82 nm (vertical) and 810.12 nm (horizontal). The different shapes and widths of the two spectra are to be expected from the material dispersion curve and the phase-matching relation, but may in part result from imperfect coupling to the multimode fibers used for spectroscopy and a nonuniform background during the measurement. For our purpose of building a heralded single-photon source, spectral indistinguishability is unimportant.}
    \label{Fig:spectrum}
\end{figure}

Historically, PPKTP crystals have mostly been used for type-I SPDC, i.e., identical polarization of the output photons, because the effective strength of the nonlinearity is lower for type-II than for type-I SPDC (for type-I, $d_{33}=10.7\,$pm/V and for type-II, $d_{32}=2.65\,$pm/V \cite{Dmitriev90a}). Nevertheless, as described previously, the signal and idler photons can be separated deterministically in the type-II case. For the second-harmonic generation (SHG), we found a conversion efficiency of about 0.03\% at 80$\,$mW pump power, which is close to the reported values in the literature. The main advantage of using PPKTP is that the poling enables collinear conversion via quasi-phase matching, which substantially improves the collection efficiency.

Figure~\ref{Fig:spectrum} shows a typical spectrum of the down-converted photons at $39^\circ$C. We used a 750$\,$mm focal length spectrometer with 600 grooves/mm grating to obtain these spectra. We can clearly see the bimodal behavior of the twin photons as we select the spectrum in polarization. We can estimate a spectral bandwidth of $\Delta\lambda=5$ and 7.5$\,$nm corresponding to values previously reported in the literature \cite{Shi04}. 

\section{Experimental results}
\label{Sec_results}
In this section, we report on our experimental results for the two coherence functions described in Sec.~\ref{Sec_Theory}. In order to measure $ g_{si}^{(2)}(\tau)$, we approximate $P_{si}(\tau)$ in Eq.~(\ref{Psit1t2}) by the rate of coincident events, $N_{si} (\tau)$, in which an idler photocount is observed at time $t$ and a signal photocount is observed in the interval $[t+\tau-\tau_{\rm coin},t+\tau+\tau_{\rm coin}]$, where $2 \tau_{\rm coin}$ is the width of our chosen coincidence window. Because of the photodetectors' time jitters, and neglecting dark counts throughout the paper, a photodetection event at time $t$ only implies the existence of one or more photons in a neighborhood around time $t$. For simplicity, we assume that the detection time corresponding to a photon that hits the detector's surface at time $t$ is uniformly distributed over the interval $[t-\tau_d,t+\tau_d]$, where $\tau_d$ is the time resolution of the photodetectors. We can then write the observed value for $N_{si} (\tau)$ in terms of $P_{si}(\tau)$ in the following way
\begin{equation}
\label{Nsi}
N_{si}(\tau)  \approx \frac{1}{2 \tau_{\rm coin}} \int_{\tau-\tau_{\rm coin}}^{\tau+\tau_{\rm coin}}{d \tau' \bar P_{si}(\tau')}  ,
\end{equation}
where
\begin{equation}
\label{barPsi}
\bar P_{si}(\tau) = \int{d t_i \int{d t_s u(t_i) u(t_s-\tau) P_{si}(t_s-t_i)}}
\end{equation}
is the coincidence rate for detecting a signal photon(s) at time $t+\tau$ and an idler photon(s) at time $t$, where $u(t) = 1/(2 \tau_d)$ if $|t|\leq \tau_d$, and zero otherwise.

\begin{figure}
    \begin{center}
        \includegraphics[width = .7\linewidth] {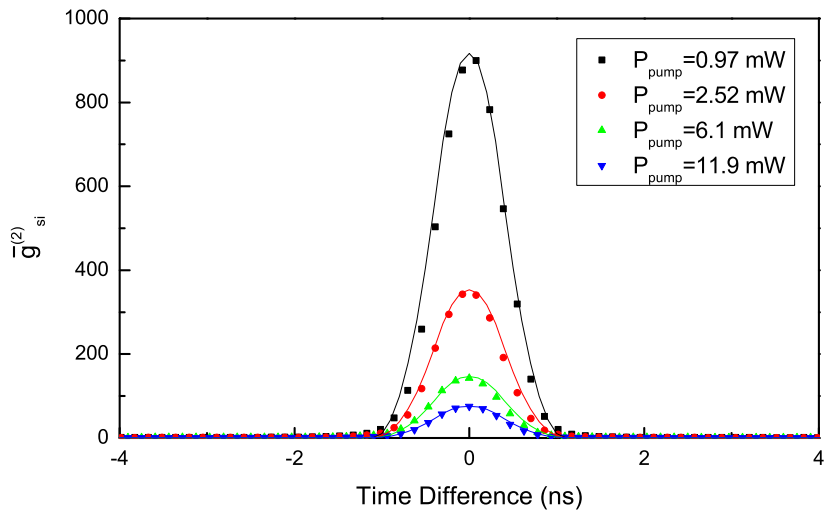}
    \end{center}
    \caption{(Color online) Measurements (symbols) and theory predictions (lines) of the time-averaged coherence function $\bar g_{si}^{(2)}(\tau)$ for the signal and idler photons at a chosen coincidence window of 0.78\unit{ns}. The low-gain regime theory curves are in striking agreement with the data using the following parameter values $R_0/P_\mathrm{pump}\approx 1.2\cdot 10^6\unit{\mbox{pairs}/(s\cdot mW)}$, $\tau_d = 350\unit{ps}$, and $B_{\rm SPDC} = 3\unit{THz}$. We measured the bandwidth by spectroscopy (see Fig.~\ref{Fig:spectrum}). The pair production rate per pump power and the detector time resolution are approximated by subjective visual fitting to the above four data sets.}
    \label{Fig:g2corr}
\end{figure}

Figure~\ref{Fig:g2corr} shows the experimental and the theoretical results for the time-averaged coherence function
\begin{equation}
\bar g_{si}^{(2)}(\tau) \equiv N_{si}(\tau)/R^2(0)
\end{equation}
for different values of pump power. Experimentally, $R^2(0)$ was determined by the product of the signal and idler count rates. For the theoretical graphs, we used the low-gain correlation functions given by Eqs.~(\ref{Rcorr}) and (\ref{Ccorr}) with $R_{\rm SPDC}=1.2$\unit{MHz} per milliwatt of pump power and $B_{\rm SPDC} = 3$\unit{THz}.
It can be seen that $g_{si}^{(2)}(0)$ drops as we increase the pump power, which is a direct result of multiple-photon contribution to the output. The peak value of $g_{si}^{(2)}(0)$ is also determined by the chosen coincidence window, here 0.78$\,$ns, because from Eqs.~(\ref{Ccorr})--(\ref{barPsi}), $N_{si}(\tau)$ has an almost fixed value for $\tau \in [-\tau_{\rm coin}+\Delta {\rm t} +\tau_d , \tau_{\rm coin}-\Delta {\rm t}-\tau_d]$, inversely proportional to $\tau_{\rm coin}$. As we get farther from the center, the time-averaged coherence function drops to its minimum value one as expected. The theoretical graphs are in striking agreement with our experimental results, which clearly demonstrate the strong temporal correlation between signal and idler beams.
\begin{figure}
    \begin{center}
        \includegraphics [width=.7\linewidth]{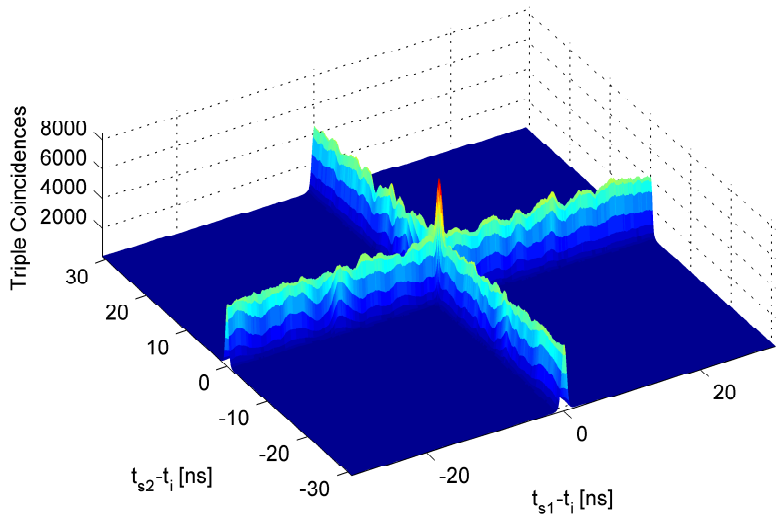}
    \end{center}
    \caption{(Color online) The triple coincidence rate of having an idler photodetection event at time $t_i$ and signal photodetection events at times $t_{s1}$ and $t_{s2}$ on, respectively, detectors $s1$ and $s2$ in Fig.~\ref{Fig:setup} for a coincidence window of $2\tau_\mathrm{coin}=0.78\unit{ns}$. Pump power for this measurement was 11.9\unit{mW}.}
    \label{Fig:triples3D}
\end{figure}

To quantify multiple-photon suppression in our HSPS, we look at $g_c^{(2)}(\tau) \equiv g_c^{(2)}(t_i, t_i+\tau|t_i) = g_c^{(2)}(0, \tau|0)$. For an ideal HSPS, we expect that $g_c^{(2)}(0)=0$. In our case, from Eq.~(\ref{gc20}), $g_c^{(2)}(0)\approx 2\cdot10^{-5} \ll 1$ at 15$\,$MHz single-photon generation rate. However, again, we are only able to measure a time-averaged version of the coherence function by approximating $P_{si}(\tau)$ with $N_{si}(\tau)$ as before and $P_{si}^{(2)}(0,\tau,0)$ with $N_{si}^{(2)}(\tau)$, the count rate for a triple coincidence of an idler photodetection event at $t_i =0$, and two signal photodetection events at $t_1 \in [-\tau_{\rm coin}, \tau_{\rm coin}]$  and $t_2 \in [\tau-\tau_{\rm coin}, \tau+\tau_{\rm coin}]$. By accounting for the resolution of the three photodetectors involved in our measurement, we obtain
\begin{equation}
N_{si}^{(2)}(\tau)  = \frac{1}{(2 \tau_{\rm coin})^2 }\int_{-\tau_{\rm coin}}^{\tau_{\rm coin}}{d t_1 \int_{\tau-\tau_{\rm coin}}^{\tau+\tau_{\rm coin}} {d t_2 \bar P_{si}^{(2)}(t_1,t_2,0)}},
\end{equation}
where
\begin{eqnarray}
\bar P_{si}^{(2)}(t_1,t_2,0) &=& \int{d t_i \int{d t_{s_1} \int{d t_{s_2} u(t_i) u(t_{s_1}-t_1)}}} \nonumber\\ &\,&  \times \, \, u(t_{s_2}-t_2) P_{si}^{(2)}(t_{s_1},t_{s_2},t_i)
\end{eqnarray}
is the multi-coincidence rate for detecting an idler photon(s) at time $0$ and two signal photons at times $t_1$ and $t_2$. Figure~\ref{Fig:triples3D} shows the experimental triple coincidence rate as a function of the two time differences. In this figure, the triple coincidence has been obtained by looking at the rate of an idler photodetection event at time $t_i$ and two signal photodetectoin events at times $t_{s1}$ and $t_{s2}$ on, respectively, detectors $s_1$ and $s_2$ in Fig.~\ref{Fig:setup}. The peak at the center of the figure represents the contribution of multiple-photon pairs, and it is proportional to $\bar P_{si}^{(2)}(0,0,0)$. The wall at $t_{s1}-t_i = 0$ in  Fig.~\ref{Fig:triples3D} represents a coincidence event between the idler photon and one of the signal photons, detected by $s1$, and it is proportional to $\bar P_{si}^{(2)}(0,\tau,0)$, where $\tau = t_{s2}-t_i$. Using Eq.~(\ref{Psit1t2ti}), one can verify that the ratio between $\bar P_{si}^{(2)}(0,0,0)$ and $\bar P_{si}^{(2)}(0,\tau,0)$ is approximately given by $(2R^2(0)+4|C(0)|^2)/(R^2(0)+|C(0)|^2) \approx 4$, where we assumed $R^2(0) \ll |C(0)|^2$ and $R(\tau) \approx C(\tau) \approx 0$. The equivalent ratio obtained from Fig.~\ref{Fig:triples3D} is about 2, which reflects the effect of time averaging in our analysis.

\begin{figure}
    \begin{center}
        \includegraphics [width=.56\textwidth]{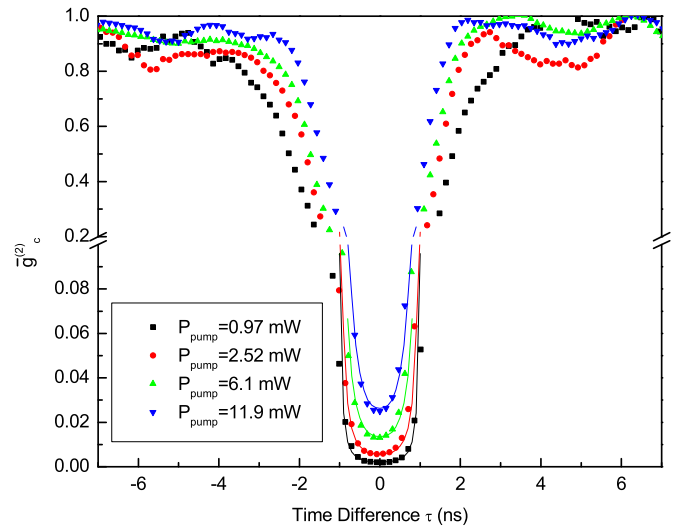}
    \end{center}
    \caption{(Color online) Measured (symbols) and calculated (lines) time-averaged conditional coherence function $\bar g_c^{(2)}(\tau)$. The theory lines were calculated using the same parameter values as in Fig.~\ref{Fig:g2corr}. The purely statistical errors of our data are on the order of the symbol size in the figure and therefore not shown. As explained in a previous article \cite{Razavi08a}, photons that are reflected twice cause the apparent ringing.}
    \label{Fig:g2dip}
\end{figure}

Figure~\ref{Fig:g2dip} shows our measurement results for the time-averaged conditional coherence function
\begin{equation}
\bar g_c^{(2)}(\tau) \equiv {N_{si}^{(2)}(\tau) R(0)}/{ [N_{si}(0) N_{si}(\tau)]}
\end{equation}
for different values of pump power, which result in different values for the observed central dip. Here, $R(0)$ is obtained from the idler count rate in the experiment. The ringing structure in Fig.~\ref{Fig:g2dip} is caused by double optical reflections \cite{Razavi08a}. The graphs, nevertheless, exhibit the signature of a good SPS as the measured value of $\bar g_c^{(2)}(0)$, at 14$\,$MHz single-photon generation rate, in Fig.~\ref{Fig:g2dip}, is below $0.03$ for $2 \tau_{\rm coin} = 0.78 \,$ns and $\tau_d = 0.35\,$ns.

\begin{figure}
    \begin{center}
        \includegraphics [width=.6\textwidth]{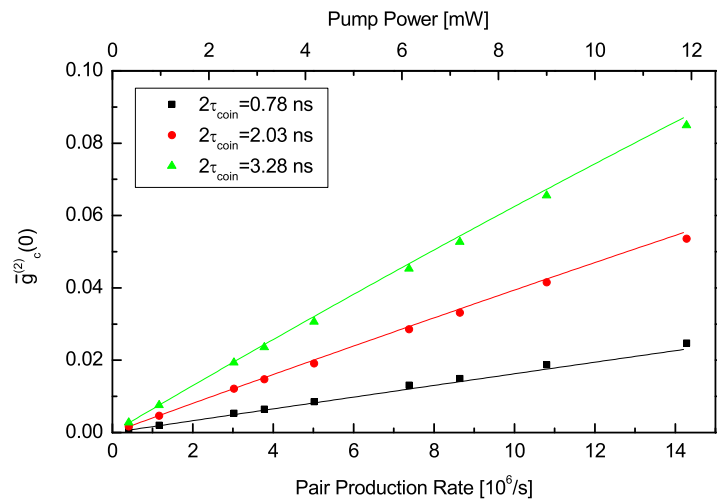}
    \end{center}
    \caption{(Color online) Measured (symbols) and calculated (lines) time-averaged conditional coherence function $\bar g_c^{(2)}(0)$ versus pump power for three different coincidence windows. The theory lines are calculated using the same parameter values as in Fig.~\ref{Fig:g2corr}.  In the low-gain regime, there is a linear increase in $\bar g_c^{(2)}(0)$ versus pump power due to the multi-photon contribution to the down-converter output.}
    \label{Fig:g2pump}
\end{figure}

By reducing the pump power we can reduce $\bar g_c^{(2)}(0)$ almost arbitrarily at the expense of reducing the total count rate. This effect has been shown in Fig.~\ref{Fig:g2pump}, where we have plotted $\bar g_c^{(2)}(0)$ versus the single-photon generation rate, $R_{\rm SPDC}$, or equivalently, the pump power. In our experiment, each milliwatt of pump power corresponds to about 1.2 million generated photon pairs per second. There is a linear growth in $\bar g_c^{(2)}(0)$ as a function of pump power, which exemplifies the contribution of multiple-photon states to the output in the low-gain regime. In this regime, from Eqs.~(\ref{Eq:gct1t2ti}) and (\ref{Rcorr})--(\ref{Eq:g_si_2}), $g_c^{(2)}(0) \approx 2/ g_{si}^{(2)}(0) \approx 2R^2(0)/|C(0)|^2 = 2 R_{\rm SPDC}/B_{\rm SPDC}$, which is proportional to the pump power. The value of $\bar g_c^{(2)}(0)$ is also a function of coincidence window as shown next.

\begin{figure}
    \begin{center}
        \includegraphics [width=.6\textwidth]{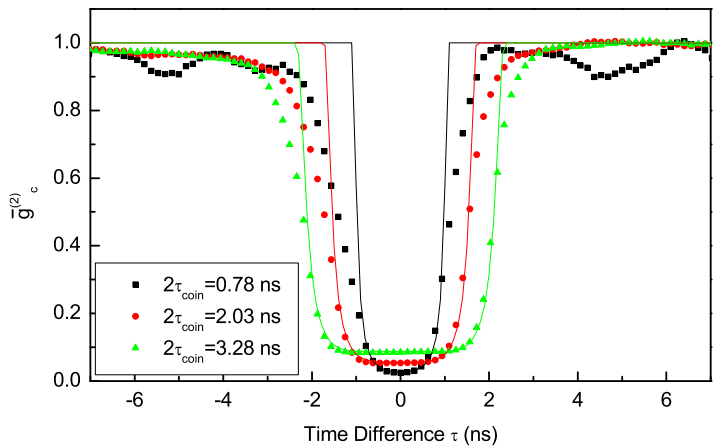}
    \end{center}
    \caption{(Color online) Measured (symbols) and calculated (lines) time-averaged conditional coherence function $\bar g_c^{(2)}(\tau)$ for three different coincidence windows. Here, the pump power was 11.9~mW corresponding to a pair production rate of approximately 14\unit{MHz}. The theory lines are calculated using the same parameter values as in Fig.~\ref{Fig:g2corr}. The theory curves only reproduce the data near the center of the dip and at very long delay times. This is to be expected because the finite time resolution and the shape of the spectrum were modeled with simplified rectangular and triangular shapes, respectively.}
    \label{Fig:g2cond_w}
\end{figure}

\begin{figure}
    \begin{center}
        \includegraphics [width=.55\linewidth]{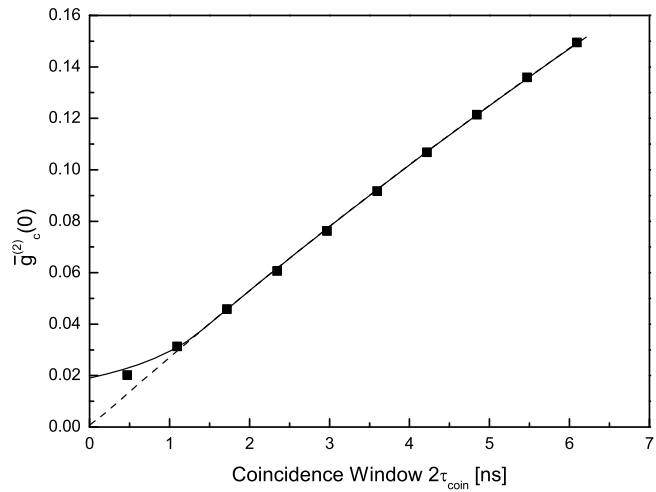}
    \end{center}
    \caption{(Color online) Experimental (symbols) and theoretical (line) results for the minimum of the time-averaged conditional coherence function, $\bar g_{c}^{(2)}(0)$, as a function of the coincidence window $2\tau_{\rm coin}$ using the same set of parameters as in Fig.~\ref{Fig:g2corr} at a pump power of 11.9\unit{mW}. The dashed line is for ideal photodetectors ($\tau_d = 0$).}
    \label{Fig:window}
\end{figure}

In Fig.~\ref{Fig:g2cond_w} one can see an example of how the conditional coherence function varies with the chosen coincidence window. Here, we have shown $\bar g_c^{(2)} (\tau)$ for three values of the coincidence window. It can be seen that the width of the central dip is almost given by $2\tau_{\rm coin}$. The measured value for $\bar g_c^{(2)} (0)$ goes down by choosing shorter coincidence windows. It does not, however, approach the actual value of $g_c^{(2)} (0)$ so long as the detector time resolution $\tau_d \gg \Delta {\rm t}$. In order to make this point clearer, in Fig.~\ref{Fig:window}, we have plotted $\bar g_c^{(2)}(0)$ versus $2\tau_{\rm coin}$. It can be seen that, for $\tau_{\rm coin} \ll \tau_d$, $\bar g_c^{(2)}(0)$ is determined by $\tau_d$, whereas, for $\tau_{\rm coin} \gg \tau_d$, it is almost linearly increasing with $\tau_{\rm coin}$. Our theoretical treatment is again well capable of reproducing the measurement results. The graph shown in Fig.~\ref{Fig:window} exemplifies the fact that a single value for $\bar g_c^{(2)}(0)$ does not bear sufficient information to quantify the source performance. At a fixed rate, the interplay between the coincidence window and the time resolution of photodetectors must also be accounted to give a proper figure of merit for an SPS. Eventually, the true value of $g_c^{(2)}(0)$ can be obtained from Eq.~(\ref{gc20}) by estimating $R(0)$ and $C(0)$. This can be done by finding the parameters that can best reproduce all or a subset of graphs shown in Figs.~\ref{Fig:g2corr}--\ref{Fig:window}.

\section{Conclusion}
\label{Sec_Conclusion}
In this paper, we theoretically and experimentally studied the coherence properties of heralded single-photon sources that use parametric down-conversion. We used the Gaussian characteristics of down-converted fields to analytically find the degree of second-order coherence between signal and idler fields as well as for the signal field, individually, when it is conditioned on the detection of an idler photon. Our theory is well capable of reproducing our experimental results, which demonstrated a high-quality source of sub-picosecond single photons. It also allowed us to study the impacts of the chosen coincidence window, the down-conversion parameters, and the resolution of photodetectors on the outcome. Such an analysis enables proper standardization of single-photon sources even with imperfect measurement devices.

\section*{Acknowledgments}
We would like to thank N. L\"utkenhaus and A. Safavi-Naeni for their technical assistance and acknowledge NSERC, CFI, ORF-RI, ORDCF, ERA, QuantumWorks, CIPI, and CIFAR for their financial support.

\section*{Appendix. $g_c^{(2)}$ calculation: discrete-mode formalism}
The analysis in this Appendix is based on looking at the system's behavior within a finite time interval or frequency band. In such cases, instead of working with the continuous-time field-operator formalism represented by $\hat E_j(t)$ and $\hat A_j(\omega)$, introduced in Sec.~\ref{Sec_Theory}, we can deal with a discrete set of annihilation operators. Here, we first develop such a multi-mode but discrete representation for the field operators in time and frequency domains. We then use our new formalism to describe the system's initial state and the measurement on the idler beam, as well as to find the post-measurement state and the conditional coherence function.

For a time interval of finite width $T$, such as $[-T/2,T/2]$, the field operator in Eq.~(\ref{Eq:field_op}) can be written as \cite{Blow90a}
\begin{equation}
e^{i \omega_j t}\hat E_j(t) = \sum_n{\hat a_{j,n} \frac {\exp[-2 i \pi n t / T]}{\sqrt{T}} }, \quad\mbox{$t \in [-T/2,T/2]$ and $j=s,i$}
\end{equation}
where
\begin{equation}
\label{Eq:ajn}
\hat a_{j,n} = \int_{-T/2}^{T/2}{dt~e^{i \omega_j t}\hat E_j(t)\frac {\exp[2 i \pi n t / T]}{\sqrt{T}} }, \quad\mbox{$j=s,i$}.
\end{equation}
Here, $\{\exp[-2 \pi i n t / T]/\sqrt{T}\}$, for integer $n$, forms an orthonormal set of basis functions that span all finite-energy functions over $t \in [-T/2,T/2]$. The operator $\hat a_{j,n}$ is the corresponding annihilation operator associated with the $n$th mode function, which represents a frequency band of effective width $2 \pi \delta f \equiv 2\pi/T$ around center frequency $2 \pi n \delta f$. Hence for $T\gg \Delta{\rm t}$, the operators $\hat a_{j,n}$, satisfying $[\hat a_{j,n},\hat a_{k,m}^\dag]=\delta_{nm}\delta_{jk}$, for $j,k\in \{s,i\}$ and integers $m$ and $n$, can describe the spectral behavior of the SPDC process with sufficient accuracy.

Alternatively, one can span the spectral field operators in Eqs.~(\ref{Eq:Bogo1}) and (\ref{Eq:Bogo2}) as follows
\begin{equation}
\label{Eq:Aj_disc}
\!\!\!\!\!\!\!\!\!\!\!\!\!\!\!\!\!\!\!\! \hat A_j(\omega+\omega_j) = \sum_n{\hat b_{j,n} \frac {\exp[ 2 i \pi n f / (2W)]}{\sqrt{2W}} }, \quad\mbox{$f \equiv \displaystyle\omega/2\pi \in [-W,W]$ and $j=s,i$},
\end{equation}
where
\begin{equation}
\hat b_{j,n} = \int_{-W}^{W}{df \hat A_j(\omega_j+ 2 \pi f)\frac {\exp[-2 i \pi n f / (2W)]}{\sqrt{2W}} }, \quad\mbox{$j=s,i$}.
\end{equation}
Here, $\hat b_{j,n}$, $j=s,i$, is the annihilation operator associated with a time interval of effective width $\delta t \equiv 1/(2W)$ centered at $n \delta t$. Again, if we choose $W$ to be much larger than $B_{\rm SPDC}$, the operators $\hat b_{j,n}$ can address the temporal behavior of the SPDC process with sufficient resolution.

Assuming $W \ge B_{\rm SPDC}$ and $TW \gg 1$, the above dual
pictures can be related to each other by plugging
Eq.~(\ref{Eq:field_op}) into Eq.~(\ref{Eq:ajn}), and then,
approximating $\int {d f \hat A_j(\omega_j + 2 \pi f) \exp(- 2 i \pi
f t)}$  by $\int_{-W}^W {d f \hat A_j(\omega_j + 2 \pi f) \exp(- 2 i
\pi f t)}$. Then, with the help of Eq.~(\ref{Eq:Aj_disc}) and some
algebra, one can obtain
\begin{eqnarray}
\label{Eq:dual1}
&\hat a_{j,n} \approx \displaystyle \sum_{m=-WT}^{WT} {\hat b_{j,m} \frac {\exp[2 i \pi n m / M]}{\sqrt{M}} }, \quad\mbox{$j=s,i$}& \\
\label{Eq:dual2}
&\hat b_{j,n} \approx \displaystyle \sum_{m=-WT}^{WT} {\hat a_{j,m} \frac {\exp[-2 i \pi n m / M]}{\sqrt{M}} }, \quad\mbox{$j=s,i$}&
\end{eqnarray}
where $M=2WT+1$, assumed to be integer, denotes the total number of modes considered for the description of the SPDC output.

The above dual formalism enables us to analytically describe the initial state of the system, the measurement performed on the idler beam and the corresponding post-measurement state of the signal beam, as well as the coherence functions of our interest. The spectral representation given by $\{\hat a_{j,n}\}$ allows us to describe $|\psi_{si}\rangle$, the state of the system at the outcome of the parametric down-converter, explicitly in the following form \begin{equation}
\label{Eq:init_st_disc}
|\psi_{si}\rangle = \bigotimes_n |\psi_{si,n}\rangle ,
\end{equation}
where, from Eqs.~(\ref{Eq:Bogo1}) and (\ref{Eq:Bogo2}),
\begin{equation}
|\psi_{si,n}\rangle = \displaystyle \sum_{k=0}^{\infty} {\frac{\nu_n^k}{|\mu_n|^{k+1}}|k\rangle_{a_{s,n}}|k\rangle_{a_{i,-n}}}
\end{equation}
is the two-mode squeezed state associated with the joint state of
harmonic oscillators represented by $\hat a_{s,n}$ and $\hat
a_{i,-n}$. In the above equation, $|k\rangle_{a_{j,n}}$ is the
$k$-photon number state associated with $\hat a_{j,n}$, $j=s,i$, and
$\mu_n \equiv \mu(2 \pi n \delta f)$, $\nu_n \equiv \nu(2 \pi n
\delta f)$.

Our calculations here mostly rely on an equivalent form of the above joint state, i.e., its Wigner characteristic function defined as follows
\begin{eqnarray}
\chi_W^{\hat a_{s,n},\hat a_{i,-n}}(\zeta_{s},\zeta_i) & \equiv & \langle \hat D (\hat a_{s,n}, \zeta_{s}) \hat D (\hat a_{i,-n}, \zeta_{i}) \rangle \nonumber \\
& = & \exp [-(|\mu_n|^2-1/2) (|\zeta_s|^2 + |\zeta_i|^2) + 2 \Re \{\mu_n \nu_n \zeta_s^\ast \zeta_i^\ast\}],
\end{eqnarray}
where $\hat D (\hat a, \zeta) \equiv \exp[\zeta \hat a^\dag - \zeta^\ast \hat a]$ is the displacement operator associated with the annihilation operator $\hat a$. The main feature of the above characteristic function is its being Gaussian with respect to its complex arguments $\zeta_{s}$ and $\zeta_i$.

By using Eq.~(\ref{Eq:dual2}), we can also find the Wigner characteristic function associated  with the state of the temporal modes of the system. The one which is of interest to us for our future calculations is
\begin{eqnarray}
\label{Eq:temp_ch}
\!\!\!\!\!\!\!\!\!\!\!\!\!\!\!\!\!\!\!  \chi_W^{\hat b_{s,k},\hat b_{s,l},\hat b_{i,0}} (\gamma_{s,k}, \gamma_{s,l},\gamma_{i,0}) & \equiv & \langle \hat D(\hat b_{s,k}, \gamma_{s,k}) \hat D(\hat b_{s,l}, \gamma_{s,l}) \hat D(\hat b_{i,0}, \gamma_{i,0})\rangle  \nonumber \\
& = & \left\langle \exp \left[ \gamma_{s,k} \displaystyle \sum_{m=-WT}^{WT} {\hat a_{s,m}^\dag \frac {\exp[2 i \pi k m / M]}{\sqrt{M}} } \right. \right. \nonumber \\
&\,&  \,\,\,\,\,\,\,\,\,\,\,\,\,  +
\gamma_{s,l} \displaystyle \sum_{m=-WT}^{WT} {\hat a_{s,m}^\dag \frac {\exp[2 i \pi l m / M]}{\sqrt{M}} } \nonumber \\
&\,& \,\,\,\,\,\,\,\,\,\,\,\,\, +
\left. \left. \gamma_{i,0} \displaystyle \sum_{m=-WT}^{WT} {\hat a_{i,m}^\dag /{\sqrt{M}} - H.c.} \right] \right\rangle \nonumber \\
& = & \exp \left[  \displaystyle -  (R_0-1/2)  (|\gamma_{s,k}|^2 +|\gamma_{s,l}|^2 + |\gamma_{i,0}|^2 ) \right. \nonumber \\
&\,& \,\,\,\,\,\,\,\,\,\,\,\,\, - 2 \Re \{ \gamma_{s,k} \gamma_{s,l}^\ast R_{k-l}\} +  2 \Re \{ \gamma_{s,k}^\ast \gamma_{i,0}^\ast C_{k}\}
\nonumber \\
&\,& \,\,\,\,\,\,\,\,\,\,\,\,\,
 \left.  + 2 \Re \{ \gamma_{s,l}^\ast \gamma_{i,0}^\ast C_{l}\}  \right], \quad\mbox{$k \neq l$ and $|k-l|<M$,}
\end{eqnarray}
where $H.c.$ denotes Hermitian conjugate,
\begin{eqnarray}
\!\!\!\!\!\!\!\!\!\!\!\!\!\!\!\!\!\!& R_n \equiv \displaystyle \sum_{m=-WT}^{WT} {\frac{1+|\nu_m|^2} {M} e^{2 i \pi n m / M}} \approx \delta_{n0} + R(n \delta t)/(2W) , \quad\mbox{$n=-W T \cdots W T$,}& \\
\!\!\!\!\!\!\!\!\!\!\!\!\!\!\!\!\!\!& C_n \equiv \displaystyle \sum_{m=-WT}^{WT} {\frac{\nu_m \mu_m} {M} e^{-2 i \pi n m / M}} \approx  C(n \delta t)/(2W),  \quad\mbox{$n=-W T \cdots W T$,}&
\end{eqnarray}
and we have assumed that $M$ is a sufficiently large prime number. The joint Wigner characteristic functions of any combination of signal and idler modes can similarly be calculated. In particular,
$\chi_W^{\hat b_{s,k},\hat b_{i,0}} (\gamma_{s,k}, \gamma_{i,0})  = \chi_W^{\hat b_{s,k},\hat b_{s,l},\hat b_{i,0}} (\gamma_{s,k}, 0 ,\gamma_{i,0})$.

The characteristic function in Eq.~(\ref{Eq:temp_ch}) has a Gaussian form and can tell us about the joint signal-idler state at different epochs of time. For instance, the joint state of $\hat b_{s,k}$ and $\hat b_{i,0}$ is entangled if and only if $C_k \neq 0$. That implies that, in the low-gain regime, modes represented by $\hat b_{s,k}$ and $\hat b_{i,0}$ are in separable states if and only if $|k| \geq \Delta {\rm t} / \delta t$. A click on the idler's photodetector at time $t_i =0$ then has only correlation with photons appearing in the signal beam during $[-\Delta {\rm t}, \Delta {\rm t}]$ interval. We clarify this issue by calculating the $g_c^{(2)}(t_1,t_2|t_i)$ below.

Without loss of generality, we assume $t_i = 0$, and, within our discrete-time formalism, we approximate $g_c^{(2)}(t_1,t_2|0)$ by
\begin{equation}
\label{Eq:gc_disc}
g_{cd}^{(2)} (k, l| 0) \equiv \frac {\langle \hat b_{s,k}^\dag \hat b_{s,l}^\dag \hat b_{s,k} \hat b_{s,l}\rangle} {\langle \hat b_{s,k}^\dag \hat b_{s,k} \rangle \langle \hat b_{s,l}^\dag \hat b_{s,l} \rangle} ,
\end{equation}
where the averaging is taken over the signal's post-measurement state, and $k$ and $l$ are integer numbers that satisfy
$t_1 \in [(k-\half) \delta t, (k+\half) \delta t)$ and $t_2 \in [(l-\half) \delta t, (l+\half) \delta t)$.

The post-measurement density operator, after a detection event on the idler mode represented by $\hat b_{i,0}$, for our three modes of interest, represented by $\hat b_{s,k}$, $\hat b_{s,l}$, and $\hat b_{i,0}$, for $k \neq l$, is given by \cite{Razavi06a}
\begin{eqnarray}
\label{Eq:rho_kl}
\!\!\!\!\!\!\!\!\!\!\!\!\!\!\! \rho_{kl}^{(pm)} = \displaystyle \frac{1} {P_{\rm{det}}} \int{\frac{d^2 \gamma_{s,k}} {\pi} \int {\frac{d^2 \gamma_{s,l}} {\pi}} \int{\frac{d^2 \gamma_{i,0}} {\pi}} \chi_W^{\hat b_{s,k},\hat b_{s,l},\hat b_{i,0}}(\gamma_{s,k},\gamma_{s,l}, \gamma_{i,0})} \times \nonumber \\
\,\,\,\,\,\,\,\,\,\,\,\,\,\,\,\,\,\,\,\,\,\,\,\,\,\,\,\,\,\,
\hat M_i \hat D(\hat b_{s,k}, -\gamma_{s,k}) \hat D(\hat b_{s,l}, -\gamma_{s,l}) \hat D(\hat b_{i,0}, -\gamma_{i,0})  \hat M_i^\dag,
\end{eqnarray}
where  $\int{ d^2 \alpha} \equiv \int_{-\infty}^\infty{d \Re{\alpha}}\int_{-\infty}^\infty{d \Im{\alpha}}$,
\begin{equation}
P_{\rm{det}} ={\langle \psi_{si}| \hat M_i^\dag \hat M_i |\psi_{si}\rangle}=1-1/R_0,
\end{equation}
and the measurement operator $\hat M_i$ is defined as follows
\begin{equation}
\hat M_i = \hat I_{b_{i,0}} - |0\rangle_{b_{i,0}\,b_{i,0}}\langle 0 | ,
\end{equation}
where $|0\rangle_{b_{i,0}}$ is the vacuum state and $\hat I_{b_{i,0}}$ is the identity operator associated with $\hat b_{i,0}$ mode. The above measurement operator accounts for one or more idler photons in an interval of width $\delta t$ around $t_i = 0$.

Similarly, the post-measurement density operator for temporal modes represented by $\hat b_{s,k}$ and $\hat b_{i,0}$ is given by
\begin{equation}
\label{Eq:rho_k}
 \!\!\!\!\!\!\!\!\!\!\!\!\!\!\!\!\!\!\!\!\!\!\!\!\!\!\!\!\!\! \rho_{k}^{(pm)} = \displaystyle \frac{1} {P_{\rm{det}}} \int{\frac{d^2 \gamma_{s,k}} {\pi} \int{\frac{d^2 \gamma_{i,0}} {\pi}} \chi_W^{\hat b_{s,k},\hat b_{i,0}}(\gamma_{s,k}, \gamma_{i,0}) \hat M_i \hat D(\hat b_{s,k}, -\gamma_{s,k}) \hat D(\hat b_{i,0}, -\gamma_{i,0})  \hat M_i^\dag}.
\end{equation}

With the help of Eqs.~(\ref{Eq:rho_kl}) and (\ref{Eq:rho_k}) and the Gaussian form of the characteristic functions, we can show that the coherence function in Eq.~(\ref{Eq:gc_disc}) is given by:
\begin{eqnarray}
\label{Eq:gckl}
g_{cd}^{(2)}(k,l|0) & = & \frac{ Q^2[Q^2 + R_0|R_{k-l}|^2 + |C_k|^2 + |C_l|^2] }{(Q^2 +|C_k|^2)(Q^2 +|C_l|^2)} \nonumber \\
&+& \frac{ 2 R_0 (R_0-1) \Re \{C_k C_l^\ast R_{k-l} \} } \nonumber \\
&-& \frac{  2 (R_0-1)|C_k|^2 |C_l|^2}{(Q^2 +|C_k|^2)(Q^2 +|C_l|^2)},
\end{eqnarray}
with $Q^2 \equiv R_0(R_0-1)^2$ and $k \neq l$, and
\begin{equation}
\label{Eq:gckk}
g_{cd}^{(2)}(k,k|0) = 2- \frac{ 2 (2-R_0)|C_k|^4}{(Q^2 +|C_k|^2)^2}.
\end{equation}
One can verify that, in the limit of $W \rightarrow \infty$, Eqs.~(\ref{Eq:gckl}) and (\ref{Eq:gckk}) converge to Eq.~(\ref{Eq:gct1t2}). This is because, in this regime, $\delta t \rightarrow 0$, hence the discrete-time annihilation operators approach to the continuous-time field operators. This proves that the heuristic approach that we employed in the previous section is indeed valid and gives us the correct result if the idler's photodetector has zero time jitter. One can also verify that the above equations reproduce all special cases we considered previously.


\section*{References}


\end{document}